\documentclass[usenatbib]{mn2e}
\usepackage{epsfig}
\usepackage{subfigure}
\usepackage{amssymb}
\usepackage{color}
\newcommand{\bn}{\begin{enumerate}}
\newcommand{\en}{\end{enumerate}}
\newcommand{\bi}{\begin{itemize}}
\newcommand{\ei}{\end{itemize}}

\def\gtorder{\mathrel{\raise.3ex\hbox{$>$}\mkern-14mu
    \lower0.6ex\hbox{$\sim$}}}
\def\ltorder{\mathrel{\raise.3ex\hbox{$<$}\mkern-14mu
    \lower0.6ex\hbox{$\sim$}}}

\newcommand{\apj}{ApJ}

\newcommand{\apjl}{ApJL}

\newcommand{\mnras}{MNRAS}

\title[What Makes the Family of Barred Disc Galaxies So Rich]
{What Makes the Family of Barred Disc Galaxies So Rich: Damping Stellar Bars in Spinning Haloes}

\author[Angela Collier, Isaac Shlosman, and Clayton Heller]
{Angela Collier$^{1}$\thanks{E-mail: angela.collier@uky.edu},
Isaac Shlosman$^{1,2}$\thanks{E-mail: shlosman@pa.uky.edu},
Clayton Heller$^{3}$
\\
$^{1}$ Department of Physics \& Astronomy, University of Kentucky, Lexington, KY 40506-0055, USA\\
$^{2}$ Theoretical Astrophysics, Graduate School of Science, Osaka University, Osaka 560-0043, Japan\\  
$^{3}$ Department of Physics, Georgia Southern University, Statesboro, GA 30460, USA \\
}

\begin{document}

\date{Accepted ?; Received ??; in original form September 29,2017}


\maketitle

\begin{abstract}
We model and analyze the secular evolution of stellar bars in spinning dark matter (DM) haloes with 
the cosmological spin $\lambda\sim 0-0.09$. Using high-resolution stellar and DM numerical 
simulations, we focus on angular momentum exchange
between stellar discs and DM haloes of various axisymmetric shapes --- spherical, oblate and prolate.
We find that stellar bars experience a diverse evolution which is guided by the ability of parent haloes
to absorb angular momentum, $J$, lost by the disc through the action of gravitational torques, resonant and
non-resonant. We confirm that dynamical bar instability is accelerated via 
resonant $J$-transfer to the halo. Our main findings relate to the long-term, secular 
evolution of disc-halo systems: with an increasing $\lambda$, bars experience less growth and
basically dissolve after they pass through vertical buckling instability.
Specifically, with increasing $\lambda$,
(1) The vertical buckling instability in stellar bars colludes with inability of the inner halo 
to absorb $J$ --- this emerges as the main factor weakening or destroying bars in 
spinning haloes; (2) Bars lose progressively less $J$, and their pattern speeds level off;
(3) Bars are smaller, and for $\lambda\gtorder 0.06$ cease their growth completely following buckling;
(4) Bars in $\lambda > 0.03$ halos have ratio of corotation-to-bar radii, $R_{\rm CR}/R_{\rm b}>2$, and
represent so-called slow bars without offset dust lanes.
We provide a quantitative analysis of $J$-transfer in disc-halo systems, and explain the
reasons for absence of growth in fast spinning haloes and its observational corollaries.
We conclude that stellar bar evolution is substantially more complex than anticipated, and bars  
are not as resilient as has been considered so far.
\end{abstract}


\section{Introduction}
\label{sec:intro}

Galactic discs are embedded in dark matter (DM) haloes of a range in the cosmological spin parameter
$\lambda\equiv J_{\rm h}/\sqrt{2} M_{\rm vir}R_{\rm vir}v_{\rm c}$, where $J_{\rm h}$ is the 
DM angular momentum, $M_{\rm vir}$ and 
$R_{\rm vir}$ --- the halo virial mass and radius, and $v_{\rm c}$ --- circular velocity at $R_{\rm
vir}$, with the mean value $\lambda = 0.035-0.04\pm 0.005$ \citep[e.g.,][]{bull01,hetz06,knebe10}. While 
discs are supported by rotation, haloes are dominated
by the random motions. When discs are embedded in DM haloes, they can serve as sources of the angular 
momentum, $J$, and haloes are perceived as sinks of $J$ \citep[e.g.,][]{sell80,debat00,atha03,marti06}. 
Hence, the angular momentum generally is expected to flow from 
the disc to the parent halo, especially when galactic bars form and facilitate the $J$-transfer. 

This description is oversimplified, because it is based on numerical simulations of nonrotating,
isolated DM haloes. Halos produced in cosmological simulations with a range of $\lambda$ usually lack 
resolution and were not analyzed
similarly. Recently, \citet{saha13} have shown that the bar instability rise time is shortened 
with increasing $\lambda$, but their analysis has been limited to the instability itself.
Furthermore, \citet{long14} demonstrated that the $J$-transfer from the disc to its
parent halo over {\it secular} time depends on $\lambda$, and its efficiency decreases sharply with increasing 
$\lambda$ --- an effect which directly opposes that of Saha \& Naab. While Long et al. have determined 
this for spherical haloes only, the importance of this effect requires a broader approach. 

Disc-halo interaction in {\it spinning} haloes has been also analysed by \citet{pete16}, which concluded
that the DM halo spin does not affect the stellar bar evolution. However, they have limited
the range of $\lambda$ to less than 0.03 and their analysis included only the first
4\,Gyr of the bar evolution. In other words, again it was limited to the time period of the bar instability itself, 
prior to the vertical buckling of stellar bars, completely 
avoiding their secular evolution.

In this paper, we demonstrate that the dependence of $J$-transfer on the cosmological spin of parent DM 
haloes over secular time is strong and a universal one, independent of the halo shape --- oblate, prolate 
or spherical. We demonstrate that stellar bar evolution is profoundly affected
by the disc-halo angular momentum transfer over wide range of $\lambda$ and time.
Furthermore, we analyze the corollaries of $J$-transfer on the evolution of galactic stellar bars.

Angular momentum redistribution in astrophysical systems is one of the main drivers of their evolution.
Gravitational torques play a major role in this process on all spatial scales, and in a broad range of
systems, from the Earth-Moon, to planetary systems, close stellar binaries, formation of compact objects, 
galaxy interactions, etc. At some instances they act on dynamical time scales, i.e., time scale comparable
with the crossing time of a system. In other cases, they act on time scales much longer than dynamical ones
--- so-called secular time scales, e.g., in accretion discs in stellar systems and compact objects. 

Any departure from axial symmetry triggers and amplifies gravitational torques. In the context
of stellar discs immersed in DM haloes, both can exhibit departures from axial symmetry.
These asymmetries can be related to the formation process of such systems, develop 
spontaneously, or as a result of interactions. 

For example, DM haloes appear universally
triaxial when forming \citep[e.g.,][]{all06,hetz06}, but tend to be axisymmetric in the contemporary
universe \citep[e.g.,][]{rix95,merri02}. This process has been demonstrated in numerical simulations with 
baryons which modify the halo shapes \citep{bere06}. 

Stellar discs can break their axial symmetry 
spontaneously \citep[e.g.,][]{hohl71,atha92a,sell93,kna95a,kna95b}, or as a result of external triggering 
\citep[e.g.,][]{holm41,toom72,nogu87,geri90}. If two gravitational
quadrupoles are present in the system, e.g., triaxial DM halo and a stellar bar, the gravitational torques
act to synchronise their rotation, by exchange of the angular momentum, although the efficiency of this 
process depends on a number of parameters.

The flow of the angular momentum in the disc-halo system has been a target of investigation for a long time.
Theoretically, it has been understood to involve resonant and non-resonant components 
\citep[e.g.,][]{lynd72,trem84,wein85}. 
Numerically, it has been detected in the first simulations involving a live DM halo \citep{sell80},
and analysed thereafter \citep[e.g.,][]{debat00,atha03,marti06,wein07a,wein07b,dubi09,villa09}.
These works have focused on $J$-transfer between barred discs and {\it non-rotating} DM haloes. In such 
systems, the halo absorbs the angular momentum, and this process involves resonant and non-resonant 
interactions between
DM and stellar orbits \citep[][]{atha03,marti06,wein07a,wein07b}. However, the exact fraction of
resonant transfer has been never measured, although \citet{dubi09} counted about 20--30\% of the halo
particles appear to be trapped in major resonances at some time of their history.

Action of gravitational torques can be described within the context of a non-local viscosity 
\citep[e.g.,][]{lars84,lin87,shlo91}, causing redistribution of mass and angular momentum in the system.
Disc stars and gas can lose or acquire angular momentum.
Stars and gas that are located inside the corotation radius, lose $J$ and move in gradually. When the gaseous 
component is present, the rate of loss of $J$
is amplified due to shocks --- unlike stars, the gas cannot reside on intersecting orbits.  Bar
formation leads to an increased central concentration in both components that lose $J$, i.e., not only in
gas but also in stars \citep[][]{dubi09}. The outer
regions of discs, outside the corotation radii, can absorb some $J$ and expand, but little mass resides 
there and so its capacity to absorb $J$ is low. In contrast, the non-rotating haloes have a large capacity 
to absorb $J$.

The evolutionary corollaries for a disc-halo system redistributing angular momentum appear to be
more obvious for the disc, which loses a non-negligible amount of $J$ and develops a bar. Beyond this
fact not much is known  --- isolated haloes have been studied mostly
non-rotating, while cosmological haloes lack numerical resolution so essential for capturing the
resonant interactions, as we have noted above. 

The most general questions that can be asked about implications for observations
of galactic bars and disc galaxy evolution can be summed as follows. Does the lifetime of the bar
depend on the spin of its parent DM halo? Does the bar strength and its pattern speed? Are the bar size
and other properties affected? 
Are there any observable effects on the shape, size, concentration, etc. of galactic discs and their 
bulges? And finally, is there a measurable effect on the halo properties, at 
least for the inner haloes? 

This paper is structured as following. Numerical aspects and initial conditions are described in \S2,
and our results of numerical modeling are presented in \S3. Next, we discuss the observational 
corollaries of our results and perform additional tests. Conclusions are given in the last section.
 
\section{Numerical techniques}
\label{sec:method}

We use the $N$-body part of the tree-particle-mesh Smoothed Particle Hydrodynamics (SPH/$N$-body) code 
GIZMO originally described in \citet{hop15}.  The units of mass 
and distance are taken as $10^{10}\,M_\odot$ and 1\,kpc, respectively. The resulting time unit
is 1\,Gyr.
We use $N_{\rm h} = 7.2\times 10^6$ particles for the DM halo, and  $N_{\rm d} = 8\times 10^5$ for stars,
in order to have mass ratio of DM particles to stellar particles of unity. Gas component is neglected 
in this work. For the convergence test, we have
doubled the number of particles to $N_{\rm h} = 1.44\times 10^7$ and  $N_{\rm d} = 1.6\times 10^6$ 
in some models. The high-resolution models resulted in a qualitatively and quantitatively similar evolution 
to the lower resolution models. The number of particles in the range of $\sim 10^6-10^7$ 
was found to be sufficient to account for resonant
interactions of stellar bar and halo orbits in disc-halo systems \citep{dubi09}.  

The gravitational softening used in the current modeling is $\epsilon_{\rm grav}=25$\,pc for stars and DM.
The opening angle $\theta$ of the tree code has been reduced from 0.5--0.7 used in cosmological simulations 
to 0.4, which increases the quality of force calculations. Our models
have been run at least for 10\,Gyr with an energy conservation of 0.05\% and angular momentum
conservation of 0.03\% over this time. 

\subsection{Initial Conditions}
\label{sec:ICs}

For the initial conditions we used the method introduced by \citet{rodio06}, see also \citet{rodio09} and
\citet{long14}, with some modifications. The basic idea of this iterative approach follows the principle that 
non-equilibrium systems will evolve in the
direction of an equilibrium. We start by generating a particle distribution with a choosen density 
distribution.

We use the standard definition of oblate and prolate ellipsoids, namely, it is oblate when $a=b>c$,
and prolate when $c>a=b$. The $c$ axis always points along the $z$ direction, and $c$ and $a$ are the polar 
and equatorial DM halo axes. Note that this definition 
includes only the axisymmetric objects, and differs from definition used by \citet{all06}, who invoked triaxial 
ellipsoids with $a>b>c$.  

In order to obtain prolate and oblate
configurations from the spherical one, we have multiplied the $z$ coordinates of particles by a factor $q=c/a$
and divided the $x$ and $y$ coordinates by $q^{1/2}$. 
This method preserves the density distribution.  To maintain consistency between the models, the 
product of principal axes, $abc$, representing the halo volume, was kept fixed.

An iteration starts by evolving the
particles from their initial positions and zero velocities for a period of 0.3\,Gyr. 
Then for each of the particles in the initial unevolved distribution, we locate the nearest evolved particle 
and copy its velocity. The directions of these updated velocities are then randomized to maintain the 
isotropic velocity dispersion. This is the end of an iteration.

Typically, about 50 iteration are required to reach 
an equilibrium which has the original density distribution and self-consistent velocities. 

To test the equilibrium, isolated haloes were evolved for 3\,Gyr, checking the invariance of the virial ratio 
of the system and its velocity dispersions.

For models with discs embedded in DM haloes, we have iterated as above in the frozen disc potential.
As the iterations do not change the halo mass profile, we have calculated the disc rotational
and dispersion velocities only once, testing if the disc remains in equilibrium after the halo
iterations. 

The disc has been constructed as a pure exponential, ignoring the bulge, and its volume density is given by

\begin{eqnarray}
\rho_{\rm d}(R,z) = \bigl(\frac{M_{\rm d}}{4\pi h^2 z_0}\bigr)\,{\rm exp}(-R/h) 
     \,{\rm sech}^2\bigl(\frac{z}{z_0}\bigr),
\end{eqnarray}
where $M_{\rm d}=6.3\times 10^{10}\,M_\odot$ is the disk mass, $h=2.85$\,kpc is its radial 
scalelength, and $z_0=0.6$\,kpc is the scaleheight. $R$ and $z$ represent the cylindrical coordinates. 

The halo density is given by \citet[][hereafter NFW]{nava96},

\begin{equation}
\rho_{\rm h}(r) = \frac{\rho_{\rm s}\,e^{-(r/r_{\rm t})^2}}{[(r+r_{\rm c})/r_{\rm s}](1+r/r_{\rm s})^2}
\end{equation}
where $\rho(r)$ is the DM density in spherical coordinates, $\rho_{\rm s}$
is the (fitting) density parameter, and $r_{\rm s}=9$\,kpc is the characteristic radius, where the power 
law slope is (approximately) equal
to $-2$, and $r_{\rm c}$ is a central density core. We used the Gaussian cutoffs at 
$r_{\rm t}=86$\,kpc for the halo and $R_{\rm t}=6h\sim 17$\,kpc
for the disc models, respectively. The halo mass is $M_{\rm h} = 6.3\times 10^{11}\,M_\odot$, and 
halo-to-disc mass ratio within $R_{\rm t}$ is 2. 

Three halo shapes have been implemented.
Spherical haloes with polar-to-equatorial axis ratios, $q=c/a=1$, oblate haloes with  $q=0.8$, and
prolate haloes with $q=1.2$.  

All DM halo models have a small flat density core of $r_{\rm c}=1.4$\,kpc for numerical reasons. 

To spin up the DM haloes, we have reversed the tangential velocities of a fraction of retrograde 
(with respect to the disc rotation) DM particles. The fraction of reversed particles
is adjusted in order to give the halo prescribed $\lambda$ value, in the range of
$0-0.09$. The implemented velocity reversals preserve the solution to the Boltzmann equation and
do not alter the DM density profile or velocity magnitudes \citep{lynd60,wein85,long14}. For axisymmetric 
haloes, the invariancy under velocity reversals is a direct
corollary of the \citet{jeans19} theorem \citep[see also][]{binn08}.

Disc radial and vertical dispersion velocities have been taken as

\begin{eqnarray}
\sigma_{\rm R}(R) = \sigma_{\rm R,0}(R){\rm exp}(-R/2h) 
\end{eqnarray}
\begin{eqnarray}
\sigma_{\rm z}(R) = \sigma_{\rm z,0}(R){\rm exp}(-R/2h) 
\end{eqnarray}
where $\sigma_{\rm R,0} = 120\,{\rm km\,s^{-1}}$ and $\sigma_{\rm z,0} = 100\,{\rm km\,s^{-1}}$. 
This leads to the global minimum in the Toomre's parameter $Q\sim 1.6$ at $R\sim 2.4h$ \citep{toom64}.
$Q$ increases toward the centre and the outer disc. 

Note, that for the purpose of clearly resolving the inner regions of stellar discs, we have constructed the
initial conditions such that long bars develop. In addition, in order to comfortably resolve the initial 
phase of the bar instability, we have decided on slightly `hotter' discs \citep[e.g.,][]{atha86}. The 
result of this choice is that the buckling instability happens slightly later in time.
These decisions, while being beneficial for the follow up analysis, do not affect the physics discussed here.

Hence the only difference between our disc-halo models are shapes of DM haloes and their spin $\lambda$.
The models have been denoted in the following way. All models are prograde with their name starting with $P$.
This letter is followed by the value of $\lambda$ multiplied by 1000, and followed by the value of $q$ 
multiplied by 10. Note, we use a capital $Q$ in the model name, not to be confused with the Toomre's
parameter. For example,  P45Q12 means a prograde model with $\lambda=0.045$, and $q=1.2$.
We define the Standard Model as that of a non-rotating spherical DM halo, P00Q10. Model P90Q12 was not 
run due to the difficulty with initial conditions.

\section{Results}
\label{sec:results}

For each $q$, all the models have identical mass distribution. Moreover, for different $q$ values,
the mass distributions are the same. All models have been evolved for 10\,Gyr. This
time scale corresponds roughly to observationally inferred, maximally uninterrupted evolution of galactic 
discs by major mergers \citep[e.g.,][]{gil02}.
Discs start axisymmetric, and develop stellar bars which evolve with time. To quantify this evolution,
we follow the bar amplitudes, $A_2$, their pattern speeds, $\Omega_{\rm b}$, and their major axes, 
$R_{\rm b}$. The bar strength has been defined as the amplitude of the Fourier $m=2$ mode,

\begin{eqnarray}
\frac{A_2}{A_0} = \frac{1}{A_0}\sum_{i=1}^{N_{\rm d}} m_{\rm i}\,e^{2i\phi_{\rm i}},
\end{eqnarray} 
where we sum over stellar particles with $R\leq 14$\,kpc, and mass 
$m=m_{\rm i}$ at azimuthal angles $\phi_{\rm i}$. The amplitude of the $m=2$ mode 
has been normalized by the monopole term $A_0$. $\Omega_{\rm b}$ is obtained
from the phase angle $\phi= 0.5\,{\rm tan^{-1}}[{\rm Im}(A_2)/{\rm Re}(A_2)]$ evolution with time.

We divide the evolution into two phases. The dynamical phase consists of the bar instability 
and terminates with the {\it first} vertical buckling instability of the bar and formation of 
boxy/peanut-shaped bulge \citep[e.g.,][]{comb90,pfen91,raha91,pats02,atha05,marti06,bere07}. 
Such bulges differ from the classical bulges which are 
supported mainly by stellar dispersion velocities, and correspond to the spheroidal component 
\citep[e.g., review by][]{korm04}. The peanut-shaped bulges have different kinematics and 
origin compared to the classical bulges.

This buckling 
weakens the bar but does not dissolve it \citep{marti04}. The weakening of the bar is dynamic
and substantial --- $A_2$ decreases sharply during this process. Recurrent bucklings act to increase 
the size of the bulge \citep{marti05,marti06}, and have other effects on the bar evolution. 
Single and double bucklings have been
observed in the models presented here (see section\,\ref{sec:buck} and Figure\,\ref{fig:a1z}). 
Following the first buckling, the bar enters its next phase, that of a secular evolution.  

\subsection{Evolution of bar amplitude in spinning haloes}
\label{sec:main}

\begin{figure*}
\centerline{
 \includegraphics[width=1.0\textwidth,angle=0] {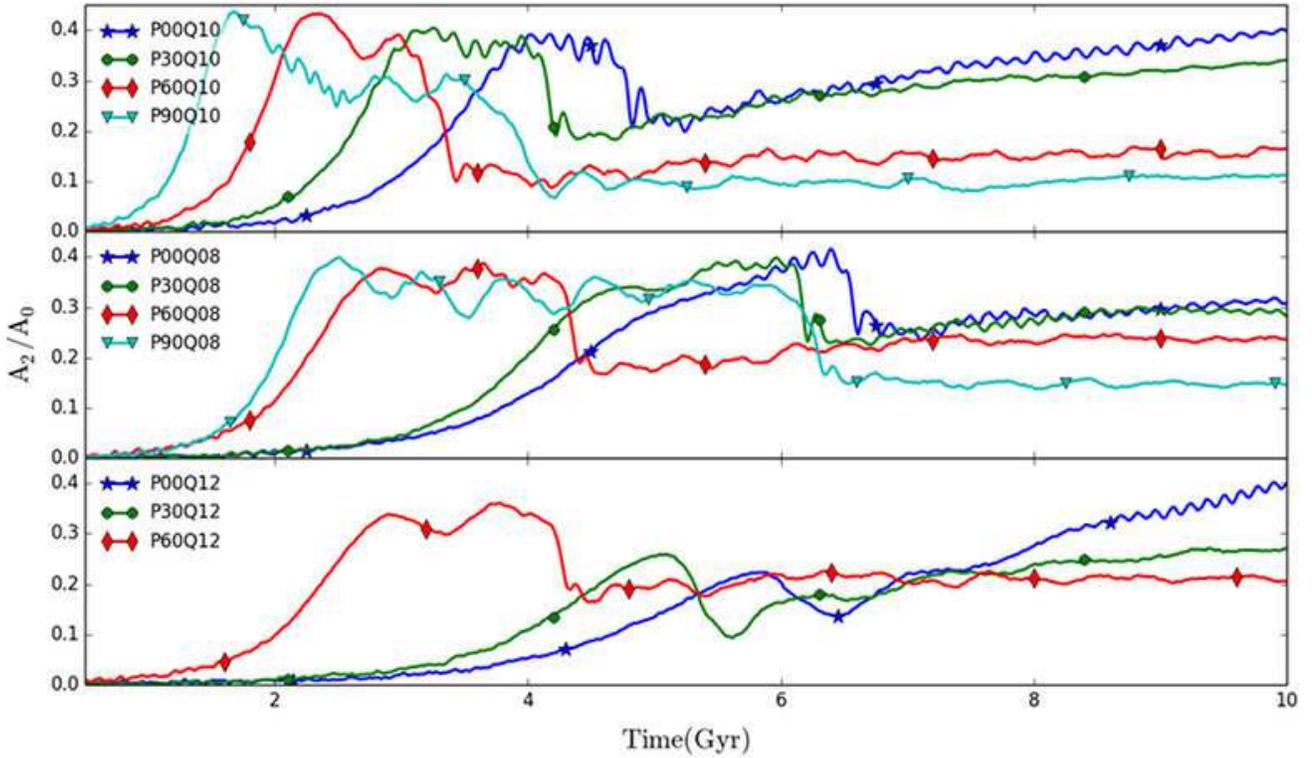}
}
\caption{Fourier amplitude $A_2$ evolution: the $\lambda$ sequence. Comparing models with the same halo 
shape for various $\lambda$. These amplitudes have been normalized by $A_0$. These colours have been 
explained in the inserts.
}
\label{fig:a2lambda}
\end{figure*}
 
\begin{figure*}
\centerline{
\includegraphics[width=1.0\textwidth,angle=0] {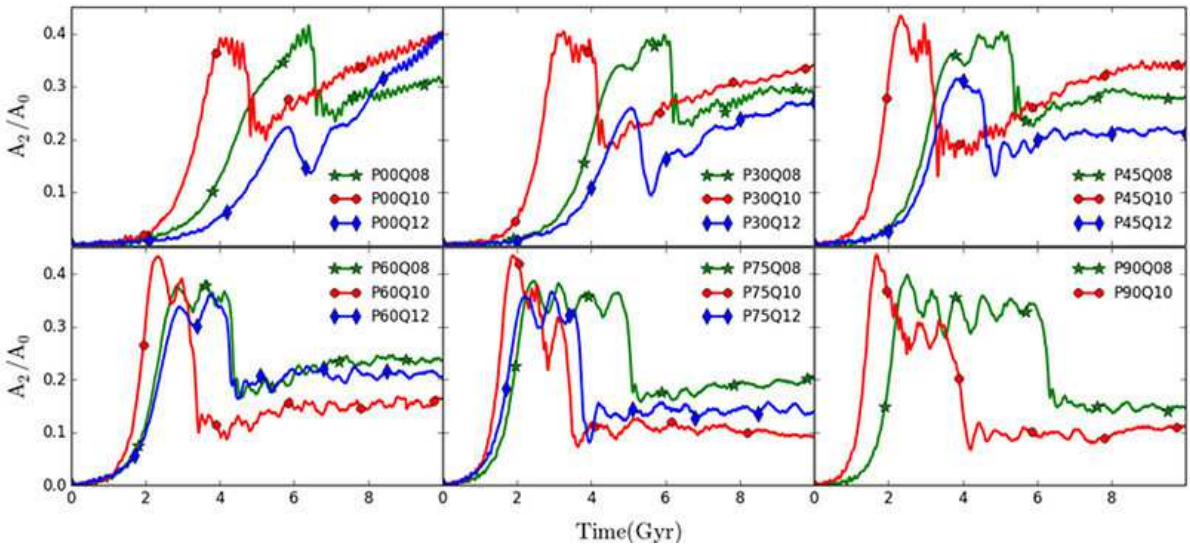}
}
\caption{Fourier amplitude $A_2$ evolution: the halo shape sequence. Comparing models with the same 
$\lambda$ for spherical, oblate, and prolate haloes separately. These amplitudes have been normalized 
by $A_0$. These colours have been explained in the inserts.
}
\label{fig:a2shapes}
\end{figure*}

Figure\,\ref{fig:a2lambda} displays the bar evolution for all models with various halo shapes and along 
the $\lambda$ sequence, while  Figure\,\ref{fig:a2shapes} focuses on a direct comparison between
different halo shapes with identical $\lambda$. Clearly, substantial differences between models 
exist along both sequences. 

First, the bar instability time scale shortens with increasing $\lambda$ for each of the halo shapes
(Figs.\,\ref{fig:a2lambda} and \ref{fig:a2shapes}), as first noted by \citet{saha13} and \citet{long14}
for spherical haloes.
The most dramatic change appears for the oblate and prolate haloes, where the bar reaches its peak at 
$t\sim 2.2$\,Gyr for $\lambda=0.09$, i.e., P90, compared to $\sim 6$\,Gyr for $\lambda=0$, P00
models. This constitutes a delay of $\sim 4$\,Gyr compared to the $\sim 2$\,Gyr for spherical models.
Hence the halo shape affects the bar instability profoundly.
 
Second, and probably of more interest, the {\it secular} growth of the bar after the first vertical buckling 
weakens with $\lambda$, for all halo shapes. Compared to the non-rotating models, those with 
$\lambda\gtorder 0.03$ display a slower growth in $A_2$ and even its leveling off at a later time.
Models with $\lambda\gtorder 0.06$ show basically no growth in $A_2$ after the first buckling. At the end
of the runs, bars in spherical haloes with $\lambda\gtorder 0.06$ exhibit the lowest amplitudes in $A_2$, 
while oblate models exhibit the highest amplitudes.  Overall, oblate, spherical and prolate haloes with larger 
$\lambda$ impede the
secular growth of the stellar bars. This conclusion confirms and strengthens that of \citet{long14}.

Third, with the exception of prolate halo models with $\lambda \ltorder 0.03$, the maximal bar amplitude 
before the first buckling is similar in all models (Fig.\,\ref{fig:a2lambda}). 

Fourth, at the $A_2$ peak, just before the first buckling, one can observe a plateau. The duration of
this plateau (i.e., its width) varies systematically among the models of each halo shape, and increases 
with $\lambda$.  

And fifth, the drop in the amplitude $A_2$, i.e., $\Delta A_2$, immediately following the first buckling 
anticorrelates with $\lambda$
for oblate and spherical models. In other words, $A_2$ after buckling reaches a deeper minimum for larger
$\lambda$. Essentially, in spinning haloes the bar nearly dissolves after buckling, with $A_2\ltorder 0.1$.
This trend is noisier for the prolate models --- still the overall
trend is clearly in tandem with other halo shape models (Fig.\,\ref{fig:a2lambda}). 

\subsection{Evolution of bar vertical buckling amplitude in spinning haloes}
\label{sec:buck}

\begin{figure*}
\centerline{
\includegraphics[width=1.15\textwidth,angle=0] {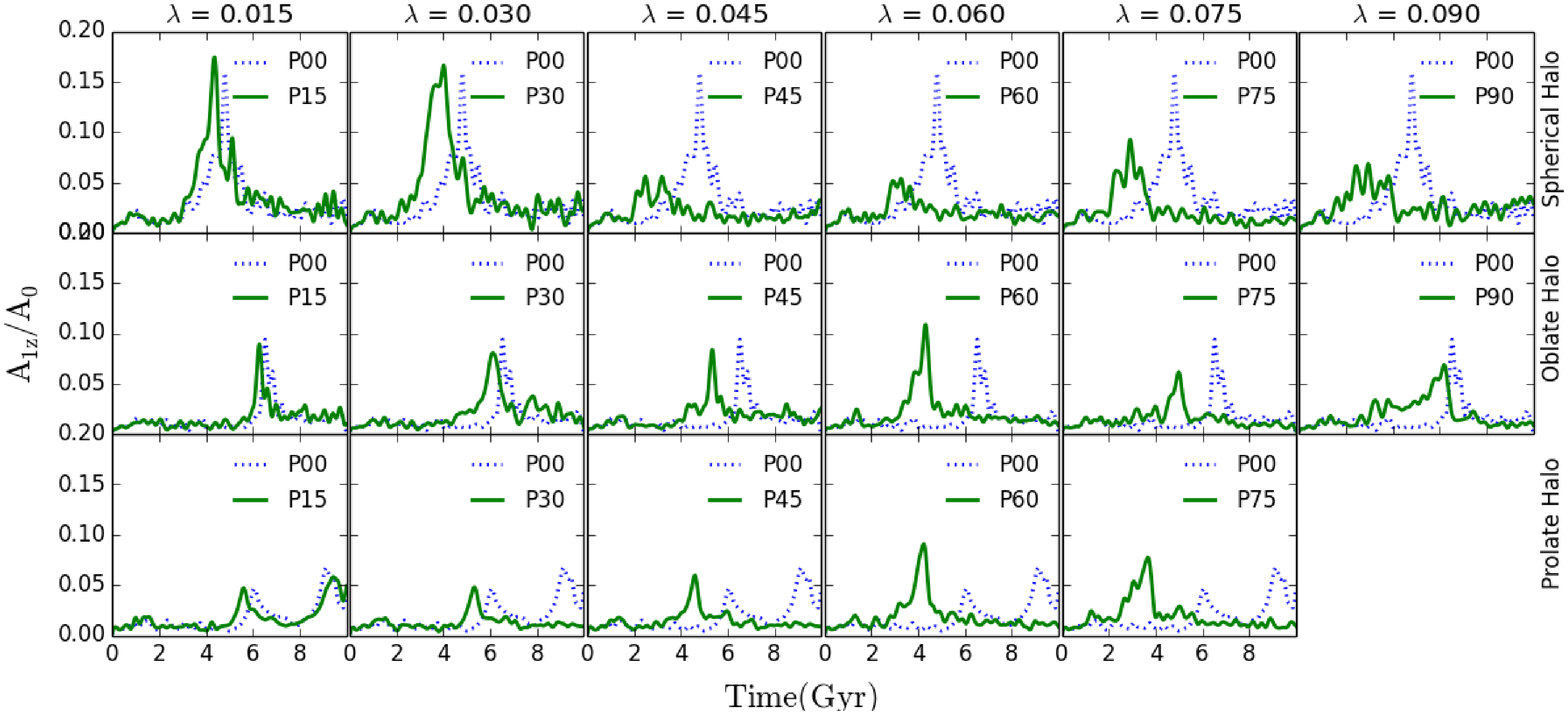}
}
\caption{Vertical buckling amplitude of stellar bars, $A_{\rm 1z}$ (green line), normalized by $A_0$, in 
spherical (top row), 
oblate (middle row) and prolate (bottom row) haloes, along the $\lambda$-sequence.  For a comparison
we superpose the buckling amplitude of $\lambda=0$ models on each $\lambda$-sequence (blue line). 
}
\label{fig:a1z}
\end{figure*}

The first vertical buckling time of stellar bars differs between the models --- the bar instability 
time scale depends
on the halo shape and its $\lambda$. The disc models are identical in all cases, so there is no
dependency on disc properties. We, therefore, take a look at the Fourier amplitude of the 
vertical buckling in these models, $A_{\rm 1z}$, 
in the $rz$-plane which is oriented along the bar major axis (Fig.\,\ref{fig:a1z}).  
We normalize this amplitude by $A_0$ calculated earlier.

Three trends can be observed here. First, the buckling happens earlier for higher $\lambda$.
Second, it happens earlier in prolate haloes, followed by the spherical and then by the oblate ones.
Third, in spherical haloes, the amplitude decreases with 
increasing $\lambda$, for $\lambda \ltorder 0.06$, then shows no preferred trend. It exhibits an opposite 
behavior in prolate models. No dependence of $A_{\rm 1z}$ maximum
on $\lambda$ is seen in oblate haloes. Lastly, $\lambda\ltorder 0.03$ prolate models experience a
double buckling, and hence exhibit two maxima in $A_{\rm 1z}$.

\subsection{Evolution of bar pattern speed in spinning haloes}
\label{sec:pattern}

\begin{figure*}
\centerline{
 \includegraphics[width=1.15\textwidth,angle=0] {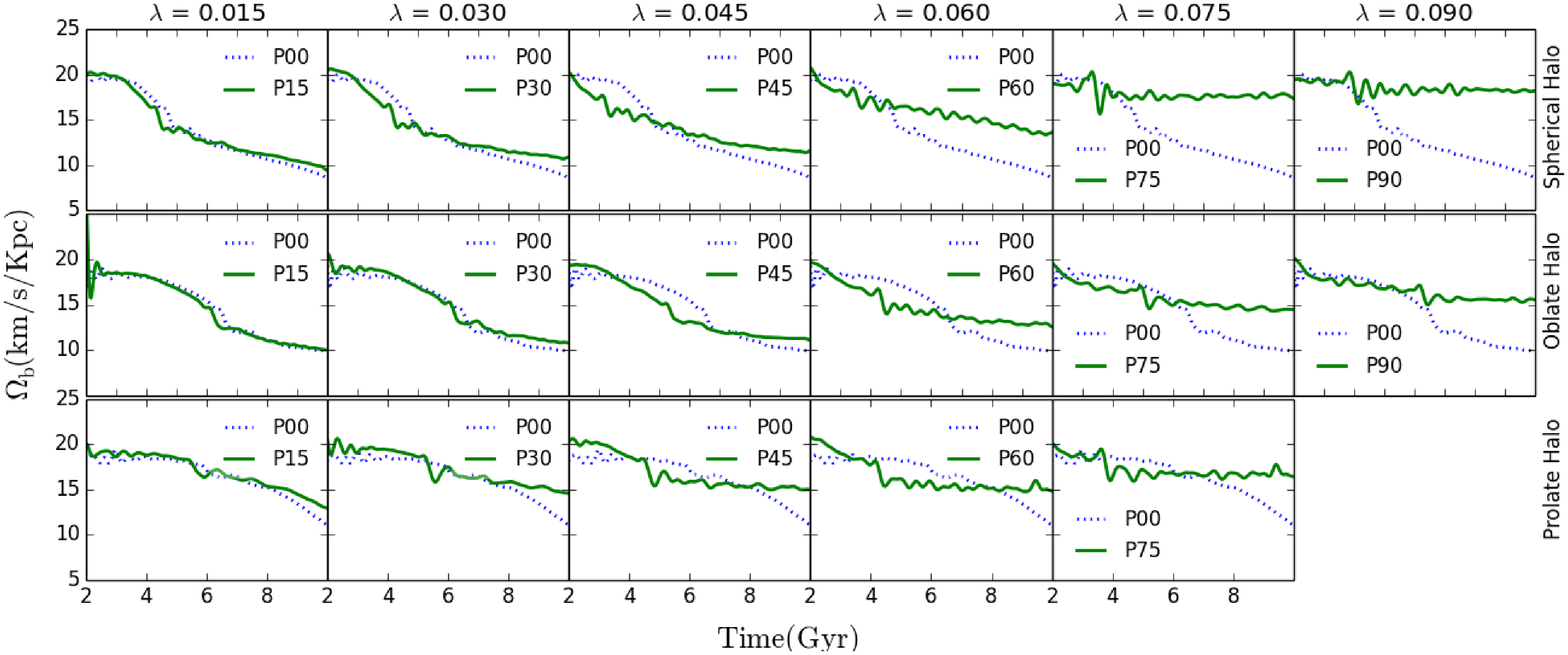}
}
\caption{Evolution of bar pattern speeds, $\Omega_{\rm b}$ (green), in spherical (top row), prolate 
(middle row) and oblate (bottom row) DM haloes and for increasing $\lambda$ (from left to right), from 
$\lambda=0.015$ to 0.09. For a comparison
we superpose the pattern speeds of $\lambda=0$ models on each $\lambda$-sequence (blue). 
}
\label{fig:omega}
\end{figure*}

Evolution of bar amplitude has a direct corollary on its rate of loss of angular momentum.
To display the kinematic properties of stellar bars in spinning haloes, we plot $\Omega_{\rm b}$
evolution in Figure\,\ref{fig:omega}. A few trends are observable here. First, the pattern speed of the
bar at the end of the simulation strongly correlates with $\lambda$. This is a consequence of the 
secular evolution of the bar, which does not regrow in amplitude after buckling in models with higher $\lambda$.
Consequently, the bar and hence the disc, lose different amounts of angular momentum in the models.

Another effect observable in Figure\,\ref{fig:omega} is that during the bar instability, {\it before} the 
buckling,
lower $\lambda$ models lose angular momentum much faster than in P00 model with a nonrotating halo.
The reason for this is that these bars grow faster in the initial stage of the bar instability.
Higher $\lambda$ models while growing faster, also buckle much earlier and their subsequent
growth is suppressed. 

Third, $\Omega_{\rm b}$ decreases abruptly during buckling for low-$\lambda$ models, while stays
flat and increases for higher $\lambda$ models. This appears to be important and we follow up
on this point in the Discussion section.

\subsection{Bar size evolution in spinning haloes}
\label{sec:size}

\begin{figure*}
\centerline{
 \includegraphics[width=1.0\textwidth,angle=0] {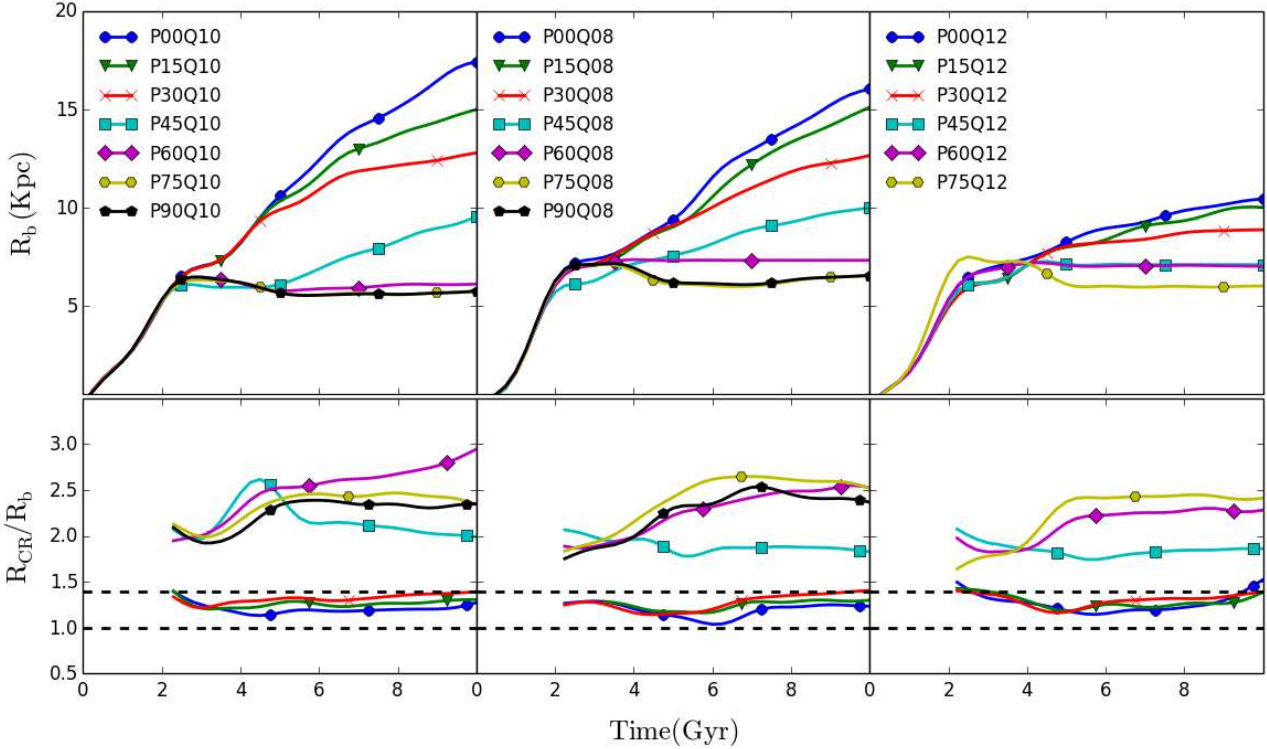}
}
\caption{{\it Top:} Bar length evolution in spherical (left), oblate (middle) and prolate (right) DM haloes.
{\it Bottom:} Bar length-to-CR radius ratio evolution within DM haloes. The dashed lines show the limits
for so-called slow bars, $R_{\rm CR}/R_{\rm b}=1.2\pm 0.2$.
}
\label{fig:ber_length}
\end{figure*}

We have determined the bar size based on the highest Jacobi energy $x_1$ orbit inside the CR \citep{marti06}.
Such orbits comprise the most important family of orbits supporting the bar density distribution. The 
$x_1$ orbits end short of the CR. 
The characteristic diagram for the main orbit families has been constructed 
\citep[e.g.,][]{cont80,Heller.Shlosman:96,bere98}, see also review by \citet{sell93}.  

Figure\,\ref{fig:ber_length} (top) shows the evolution of $R_{\rm b}$. The longest bars reside in the
spherical haloes by $t=10$\,Gyr, but evolution of bars in oblate haloes is very similar. The growth of
bars in the prolate haloes is very slow after buckling. Bars in
P00 models grow longest and their growth is fastest and monotonic, with an inflection around the
time of vertical buckling. For $\lambda\gtorder 0.06$, bars do not grow at all in all models after
buckling. 

We have also measured the ratio $R_{\rm CR}/R_{\rm b}$ (Fig.\,\ref{fig:ber_length}, bottom). Bars that extend
to the vicinity of the CR, have a narrow range of $R_{\rm b}/R_{\rm CR}\sim 1.2\pm 0.2$, so-called fast bars, 
while those that fall short of CR are slow bars \citep[e.g.,][]{teub85,atha92b}.
This result has been confirmed in \citet{marti06}, and we reproduce it here for models with $\lambda\ltorder
0.03$ for spherical and oblate haloes. For larger $\lambda$, this ratio lies outside the $1.2\pm 0.2$ range
for the entire time of their evolution. It is also true for prolate haloes with any spin. These bars, therefore,
are slow bars, and end well before the CR.

\subsection{Angular momentum transfer in oblate, spherical and prolate haloes}
\label{sec:Jdot}

\begin{figure*}
\centerline{
 \includegraphics[width=1.0\textwidth,angle=0] {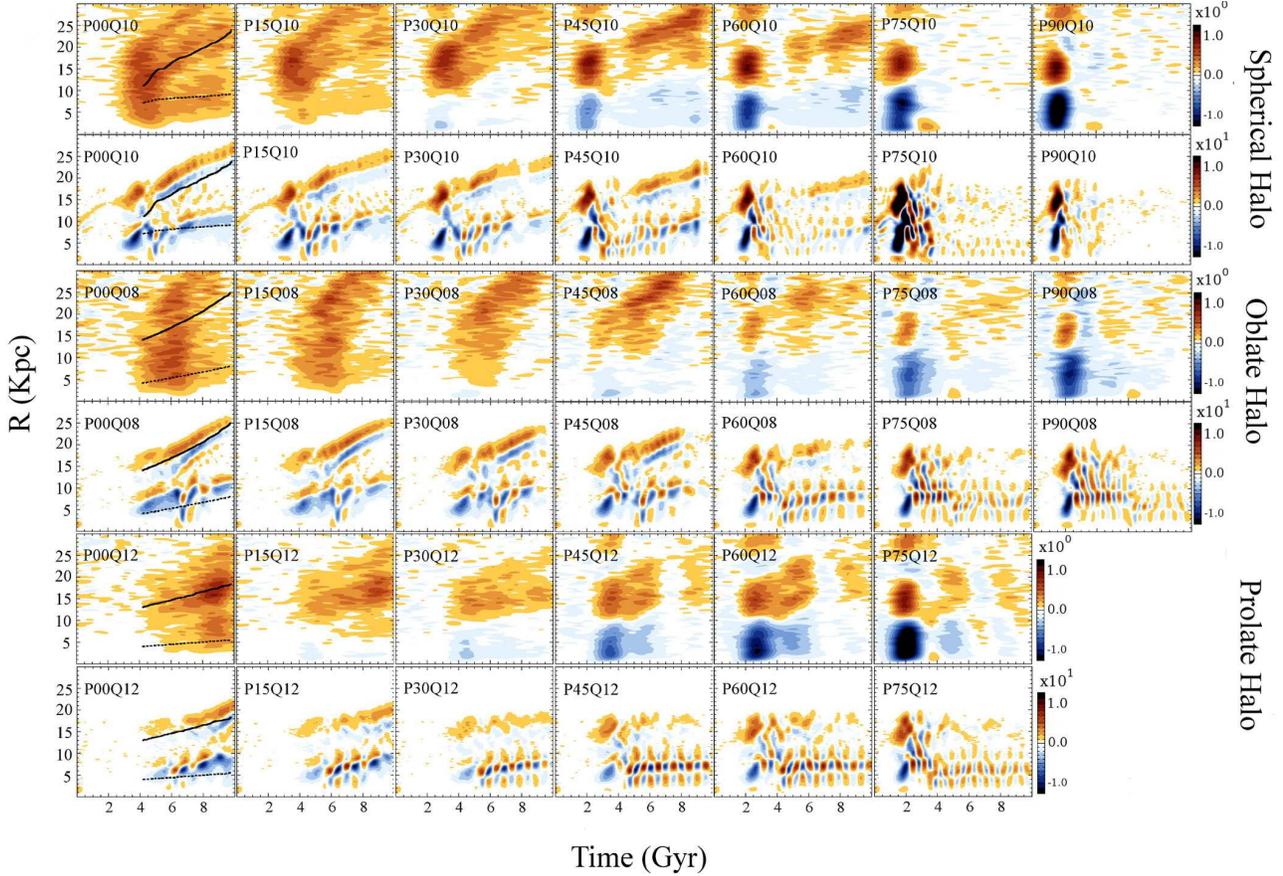}
}
\caption{Rates of angular momentum flow, $\dot J$, as a function of a cylindrical radius and time 
along the $\lambda$ sequence, for spherical (top two rows), oblate (middle two rows) and prolate
(bottom two rows) haloes and the embedded stellar discs.
The colour palette corresponds to gain/loss rates (i.e., red/blue) 
using a logarithmic scale in colour. The cylindrical shells $\Delta R = 1$\,kpc extend to  $z=\pm 3$\,kpc,
both for the haloes and their discs. Positions of major linear resonances in the disc, ILR and CR, have been 
delineated by solid and dashed lines in P00 models. 
}
\label{fig:Jdotmap}
\end{figure*}

Next, we quantify the angular momentum flow in the disc-halo systems which develop stellar bars.
In this, we follow the method developed by \citet{villa09} and \citet{long14}. This method
tracks the total angular momentum rate transfer between the disc and the DM halo, i.e., resonant
and non-resonant ones \citep{atha03,marti06}, but also can reveal the flow between various disc and
halo radii. For this purpose, 
we divide the disc and its host halo into nested 
cylindrical shells. Then construct a two-dimensional map  of the rate of $J$ change in each shell 
as a function of $R$ and $t$. the resulting colour-coded maps are shown for spherical, oblate and prolate 
(Fig.\,\ref{fig:Jdotmap}) haloes, for the $\lambda$-sequence,  and the associated stellar discs. 
The top row in each Figure exhibits the rate of angular momentum flow in DM haloes, 
$\langle\dot J_{\rm DM}\rangle \equiv (\partial J_{\rm DM}/\partial t)_{\rm R}$, while the bottom row,
shows the rate of the $J$ flow in the stellar discs, $\langle\dot J_*\rangle 
\equiv (\partial J_*/\partial t)_{\rm R}$. The brackets indicate the time averaging at $R$. 

The colours in the above Figure represent the absorption/emission (red/blue) of the angular
momentum by the DM (top) and disc (bottom) material. The colour palette has been normalised 
the same way for all discs and (separately) for all haloes. The continuity of these colours represent
the emission/absorption of $J$ by the main resonances in the DM haloes and stellar discs, as well
as the non-resonant contribution.

The evolution of linear resonances is shown by
continuous lines for $\lambda=0$ models only. For example, the emission of $J$ by the inner Lindblad 
Resonance (ILR) in the
disc follows the lower blue band drifting to larger $R$ with time in model P00Q10 
(Fig.\,\ref{fig:Jdotmap}, lower left frame). The additional blue band corresponds to the 
Ultra-Harmonic Resonance (UHR). The dominant red band follows the CR and the Outer Lindblad 
Resonance (OLR). 

This Figure is divided into three pairs of horizontal rows representing haloes (top) and discs (bottom),
each, for spherical, oblate and prolate haloes.
The upper left frame, showing the P00Q10 model, exhibits only absorption (red) by
a halo with no or low net angular momentum. However, moving along the $\lambda$ sequence, we observe
profound differences in the absorption/emission of $J$ by both the disc and the halo.

First, we invoke the Standard Model P00Q10 in order to understand the colour palette. 
The upper frame of Figure\,\ref{fig:Jdotmap} displays an intense absorption by the DM halo
after $\sim 3$\,Gyr. This corresponds to the bar strength $A_2 \gtorder 0.2$ in the 
Figure\,\ref{fig:a2lambda}, for this model. The main region in the halo which participates in this
$J$ absorption is within $\ltorder 10$\,kpc. So once the bar acquires non-linear amplitude, it
facilitates the $J$ transfer to the halo. 

For the model P00Q30, this happens earlier, at $\sim 2$\,Gyr. And, what is important, two halo
regions participate in the $J$ transfer now: a weak {\it emission} inside 10\,kpc, and absorption 
between 10--20\,kpc. Moving to larger $\lambda$, the inner region of
the halo emits $J$, while the outer one absorbs it. The absorption strength stays the same
when advancing to P90Q10 while the emission strengthens substantially. 

When comparing the maxima of $J$ absorption by the halo with the approximate positions of the main 
resonances,
we observe that the main region between the ILR and OLR dominates the process in the P00Q10.
as we move along the $\lambda$ sequence, the absorption by the ILR disappears and is reversed
to emission, while that of the
OLR increases. For $\lambda > 0.03$, the ILR starts to emit $J$. while the absorption is dominated
by the CR--OLR region.

\begin{table*}
\caption{Fractional change in the angular momenta of DM haloes from $t=0$ to $t=10$\,Gyr for spinning models 
with increasing $\lambda$.}
	\begin{tabular}{|c||c||c||c||c||c||c|}
		 \hline
		 \multicolumn{7}{|c|}{Halo $\Delta\,J/J(t=0)$} \\
		 \hline
		 Halo Shape&P15&P30&P45&P60&P75&P90\\
		 \hline
		 \hline
		 Spherical&0.1453&0.0633&0.0220&0.0114&0.0047&0.0048\\
		 \hline
		 Oblate&	0.1373&	0.0525&	0.0282&	0.0115&	0.0106&	0.0045\\
		 \hline
		 Prolate&	0.0917&	0.0234&	0.0099&	0.0041&	0.0035& \\
		 \hline
	\end{tabular}
\label{table:deltaJtable}
\end{table*}

In order to demonstrate the efficiency of angular momentum absorption by the DM halo, we have
calculated the fractional increase in $J$ for all haloes that have non-zero spin at $t=0$. 
Table\,\ref{table:deltaJtable}
shows the change in $J$ over the simulation time, $\Delta J=J(t=1-\,{\rm Gyr})-J(t=0)$, normalised
by $J(t=0)$. Clearly, along the $\lambda$ sequences for various halo shapes, this ratio is decreasing.
The decrease is significant, e.g., the P90 model haloes acquire about 30 times less angular momentum,
compared to P15 models. Hence, the efficiency of $J$ absorption by the haloes along $\lambda$ sequence
decreases.

\section{Discussion}
\label{sec:discuss}

We have analyzed evolution of stellar bars in galaxies with spinning DM haloes, with the cosmological 
spin $\lambda\sim 0 - 0.09$, which encompasses all the expected range. Various axisymmetric halo
shapes have been invoked, namely, oblate, spherical and prolate. We focus on secular evolution of
stellar bars under these conditions, and discuss implications for disc evolution. 

Our main result is that spinning haloes profoundly affect the bar properties,
which was not taken into account so far when addressing galaxy evolution. It was shown recently
that the bar
instability in axisymmetric disks is accelerated and so is the bar growth during this
dynamical phase, i.e., before they reach the maximum strength given by $A_2$ \citep{saha13,long14}.
Our main finding is that 
after bars experience vertical buckling instability, their strength decreases sharply. This decrease is
more dramatic for larger $\lambda$. Essentially, bars are dissolved for $\lambda > 0.06$, leaving
a weak oval distortion behind.

Second, in the subsequent secular phase of evolution, the bar growth, in strength and in size, is 
severely curtailed 
with increasing $\lambda$. For $\lambda\gtorder 0.06$, bar growth is completely damped and $A_2$
remains flat. Next,
for $\lambda\ltorder 0.03$, bars  extend to near CR, i.e, the ratio of $R_{\rm CR}/R_{\rm b}\sim 
1.2\pm 0.2$ remains in the narrow range (i.e., fast so-called bars). 
For higher $\lambda$, this ratio is substantially larger than 1.4, offset dust lanes are not expected
and the bars are defined as slow. 

Finally, the rate of angular momentum flow from the disc to the DM halo
decreases along the $\lambda$ sequence, after the buckling phase, with $J$ transfer going both ways
as shown by the $J$ flow maps. A clear indication of this process is the temporal speed up of the bar 
tumbling at the end of the buckling instability in higher $\lambda$ models.
This behavior has substantial corollaries to the bar growth --- unable to lose its $J$ or even increasing
it, the bar amplitude is damped even more, and its pattern speed stops to decrease. 

The behaviour of the bar amplitude, $A_2$, during the bar instability along the $\lambda$ sequence has 
been analysed by \citet{saha13}, prior to buckling only, and by \citet{long14}. The angular momentum transfer
from the disc to the halo is amplified due to the increase of the fraction of prograde orbits in 
the halo which are capable to resonate with the disc orbits. The subsequent secular evolution
that has been reported by  \citet{long14} is confirmed and further analysed in the present work.

What processes accompany the buckling of stellar bars and their subsequent evolution in spinning haloes? 
We start by focusing on the $J$ redistribution in our models (Fig.\,\ref{fig:Jdotmap}). The low-$\lambda$ 
models, P00 and P15
in all halo shapes, show a pure absorption of $J$ by the DM halos. This absorption is complemented 
by a strong emission of $J$ by the embedded discs, mostly by their ILRs. Some of this emission is
absorbed by the outer disc, but this weakens with time. 

However, for higher $\lambda$ models, emission of $J$ by haloes appears and strengthens with an increasing 
{\it inner} 
halo spin, becoming very strong. At the same time, absorption of $J$ by the haloes shifts gradually to larger $R$,
and the $J$ transfer essentially disappears soon after the buckling. By the end of the buckling, discs
exhibit strong absorption in the CR-OLR region, in all models. 
Hence, we conclude that the $J$-transfer goes from the disc to the halo in low-$\lambda$ models, and
both ways for haloes with higher spin. 

Additional argument in favour of disc receiving $J$ from the halo can be made by analysing $\Omega_{\rm b}$
behaviour during buckling (\S\ref{sec:pattern}). For lower $\lambda$, 
we observe a flattening and a subsequent drop in $\Omega_{\rm b}$, corresponding to the slowdown of the bar,
while for larger $\lambda$, we see an increase  in $\Omega_{\rm b}$, corresponding to a sudden
speedup of the bar (Fig.\,\ref{fig:omega}). Again, this behavior of $\Omega_{\rm b}$ is similar for all halo 
shapes.
In the absence of the gas component, the only source of $J$ under these circumstances is the inner halo.

The angular momentum received by the bar is not only deposited in the tumbling of the bar, but also
in the increase of the inner circulation within the bar.  Bars that lose $J$, slowdown {\it and} become 
stronger, because loss of circulation leads to increasingly radial stellar orbits within the bar. 
Similarly, with increase
of the internal circulation, the orbits become more circular, and the bar weakens. This additional
weakening of the bar, i.e., of its $A_2$, contributes to the larger drop in $A_2$ with increasing
$\lambda$  --- the bar receives larger amount of $J$ in haloes with higher spin.

Higher $\lambda$ haloes in Figure\,\ref{fig:Jdotmap} also transfer some of $J$ to larger radii, i.e., `talk 
to themselves.'  Note, that for spinning DM haloes with no discs, such a behavior has been predicted by  
\citet{papa91}, based on theoretical analysis. Lower $m$ modes have been stated to be responsible for this
evolution. We have followed the development of $m=2$ modes in the parent DM haloes as well, but unlike
the Papaloizou et al. models, this process is controlled by the non-axisymmetric modes in the {\it disc}. 
We discuss these modes elsewhere.

As can be seen in Figure\,\ref{fig:Jdotmap}, the angular momentum transfer between the disc and its
halo essentially cease after buckling for larger $\lambda$. The oval distortion which remains
in the disk does not grow in amplitude $A_2$, and so the bar does not reform. We address this issue in
the next section.

\subsection{Spinup DM halo and bar damping}
\label{sec:spinup}

\begin{figure}
\centerline{
\includegraphics[width=0.47\textwidth,angle=0] {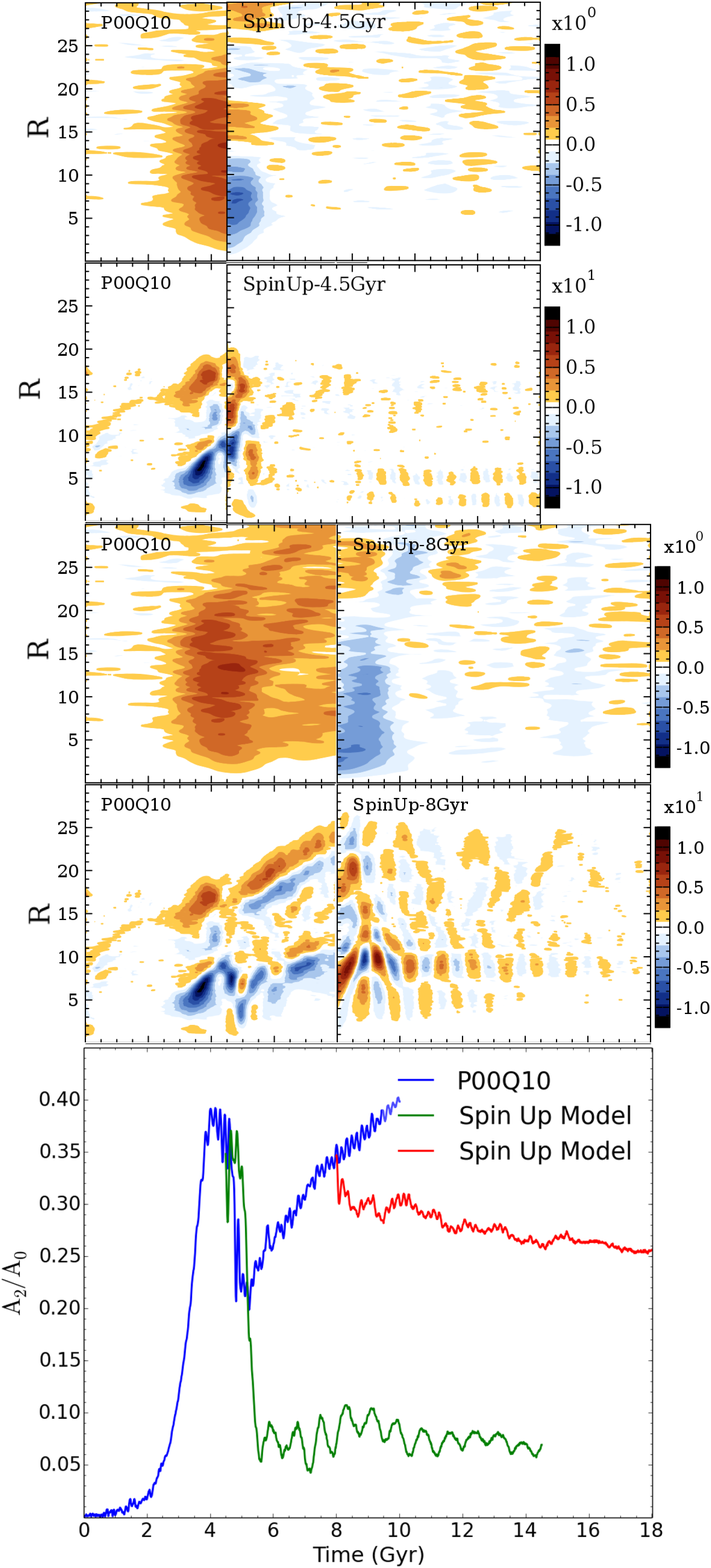}
}
\caption{Spinup of DM halo for P00Q10 model to P90Q10 experiments at $t=4.5$\,Gyr and 8\,Gyr. 
({\it a)}. Rates of angular momentum flow $\dot J$,
as in Figure\,\ref{fig:Jdotmap} but before and after the spinup at 4.5\,Gyr (top two rows). The 
left columns correspond to $J$-flow in the P00Q10 model prior to the halo spinup. The right column 
displays the $J$-flow after the spinup. The top row shows the emission and absorption of $J$ by the 
halo, while the bottom row shows the same for the disc. ({\it b}). Same as in $a$ but with the
spinup at $t=8$\,Gyr. ({\it c}). Evolution of $A_2$ amplitudes, before and after the spinup, and
comparison with the P00Q10 model. 
}
\label{fig:Jdotmap_spinup}
\end{figure}

The central question, is why will the bar not reform after the buckling for a range in $\lambda$?
First, we confirm that this is a 
robust behaviour and not a numerical fluke, and perform additional experiments.

The DM halo has been found to be a recipient of the angular momentum from the barred disc, as discussed
in \S\ref{sec:intro}. Numerical simulations have determined that this angular momentum 
transfer from the disc to DM halo involves lower resonances which trap disc and halo particles and
amplify their interactions \citep{atha03,marti06,wein07a,wein07b,dubi09}. Furthermore, \citet{villa09} 
have argued in favour of the dual role played by the DM haloes. Namely, more massive haloes within the
disc radius weaken the dynamical bar instability, while facilitating the secular growth of the bar. In all 
these works, the analysis has been limited to non-rotating haloes, mostly of a spherical shape,
with rare  exceptions \citep[e.g.,][]{bere06,atha13}.

\citet{saha13} and \citet{long14} have shown that the bar instability time scale shortens
with $\lambda$. Finally, \citet{long14} have demonstrated that faster spinning haloes damp the amplitude
of stellar bars during their secular evolution in spherical haloes. 
Here we have confirmed these previous works and have shown that the dynamical and secular evolution of
bars indeed depend on the cosmological spin parameter of their parent DM haloes.

To confirm that halo angular momentum plays the crucial role in damping stellar bars in spinning haloes,
we have performed a number of numerical experiments described below. In the 
first set of experiments, we have used the spherical
non-rotating halo in P00Q10 at $t=8$\,Gyr and 4.5\,Gyr, and spun it up to $\lambda\sim 0.09$, i.e., to 
the halo in 
P90Q10. This has been performed using the method described in \S\ref{sec:ICs}. In a 
second set of experiments, shown in the next section, we used the spherical, fast spinning halo in 
P90Q10, and spun it down to $\lambda = 0$, using the same method (\S\ref{sec:spindown}).

Figure\,\ref{fig:Jdotmap_spinup} displays the bar amplitude evolution (bottom frame) before and after the 
halo spinup at $t=8$\,Gyr and at 4.5\,Gyr. We have run these models for an additional 10\,Gyr, to test 
their behaviour. Prior to spinup, the stellar bar had been growing secularly, i.e., P00Q10 model, almost 
reaching its 
pre-buckling values of $A_2$. After the spinup at $t=8$\,Gyr, it stopped strengthening and even started 
a moderate decay. The middle-top frame
displays the rate and direction of the $J$ flow before and after the spinup. Prior to the spinup, the halo 
had been only absorbing $J$. After the spinup, it started to emit $J$, except in the region of $\gtorder
15$\,kpc, which still shows some absorption. If the halo is unable to absorb, the bar cannot grow,
and this is exactly what we detect. After $\gtorder 10$\,Gyr, there is basically no exchange of angular
momentum in the system.

Prior to the spinup, the disc had been emitting $J$ mainly at its ILR, and switched to absorption after 
the spinup. For the next 2\,Gyr, we observe  $J$ flow
from the parent halo to the disc. Subsequently, the halo and the disc are not engaged in the $J$ transfer.  

Therefore, the spinup of the halo
resulted in the angular momentum transfer to the disc for a period of about 2\,Gyr, followed by a complete
cessation of $J$-transfer between the two morphological components, {\it despite existence of a moderate
strength bar}.  

\begin{figure}
\centerline{
\includegraphics[width=0.43\textwidth,angle=0] {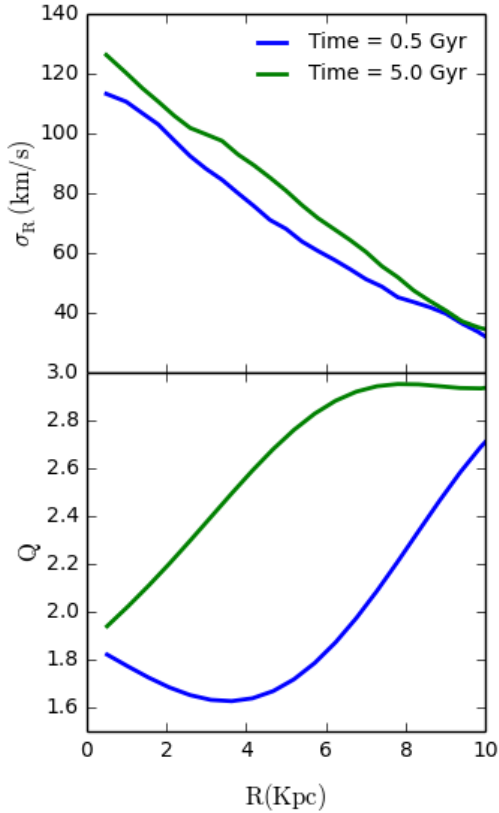}
}
\caption{Comparison between $\sigma_{\rm R}$ (top) and Toomre's $Q$ (bottom) for P90Q10 model at
$t=0.5$\,Gyr and at $t=5$\,Gyr, following bar vertical buckling and near dissolution. Note that
for the former time, the bar did not form yet, and for the latter one, it has dissolved. Hence,
this Figure shows either initial $\sigma_{\rm R}$ or dispersion velocities of de-correlated orbits in the 
disc.
}
\label{fig:P90_sigmaR_Q}
\end{figure}

\begin{figure}
\centerline{
\includegraphics[width=0.47\textwidth,angle=0] {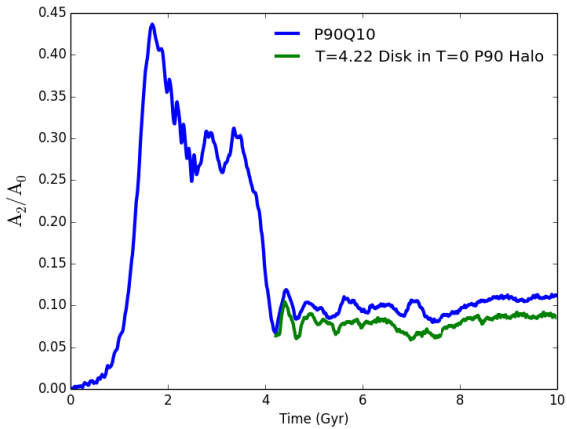}
}
\caption{Test: $A_2$ for P90Q10 model --- replacing the DM halo at $t=4.22$\,Gyr. After buckling of the bar 
the DM halo is replaced by the same halo at $t=0$. Note that the bar instability is completely suppressed
and the $A_2<0.1$ and remains flat.
}
\label{fig:A2P90sigma_test}
\end{figure}

To further test the bar evolution in spinning haloes, we have repeated the spinup of the DM halo in P00Q10 
model at $t\sim
4.5$\,Gyr (Fig.\,\ref{fig:Jdotmap_spinup}, top two rows ). The main difference with the previous experiment 
lies in that the spinup happens as the buckling develops. Indeed, the subsequent evolution of 
the system differs
profoundly from the previous experiment --- the bar is nearly completely dissolved within $\sim 1$\,Gyr from
the spinup. Thus it mimics the evolution of P90Q10 model. 

The explanation to this interesting behavior is related to the orbital evolution in the disc during the
buckling instability. \citet{marti04} have shown that the outer part of the bar, beyond the ILR, is
dissolved in the buckling, due to the increase of the fraction of chaotic orbits there, as shown
by the surface of sections. With an increase of the fraction of chaotic orbits, the area of the regular
orbits decreases and the invariant curves which enclose the region of chaotic orbits start to dissolve. 
Chaotic orbits thus `leak' through the invariant curves at Jacobi energies above the
ILR.  The bar shortens,
but survives and quickly regains its strength by transferring its angular momentum to the parent halo. 
Thus, the bar survives the buckling, but this statement is limited to nonrotating haloes.

The tandem of buckling instability and spinning haloes leads to a different outcome --- the bar amplitude
declines more than in the nonrotating haloes, because the combination of the spunup halo and the
buckling result in additional decline in $A_2$, as discussed above. Dissolution of the bar populates
the disc with orbits with large radial dispersion velocities. These orbits have been confined by the
bar before the buckling --- now they are de-correlated in the absence of the bar.

The relationship between bar dissolution and the fraction of chaotic orbits had been first discussed in
the context of the bar strength \citep{teub83,teub85,atha83}. For example, bars with axial ratios larger 
than 5:1 should
dissolve as they are dominated by chaotic orbits. Chaotic orbits will diffuse in the phase space
being limited only by energy conservation. In other words, they will de-correlate, leading to the bar
washing out. So one should expect that such de-correlated orbits from dissolved or even nearly dissolved
bars will contribute to larger radial dispersion velocities in the disc.

Returning to the problem at hand, we reiterate the question:
what prevents the bar from reforming after buckling in spinning haloes? After all, the disc becomes 
nearly axisymmetric and the halo is identical to that of P90Q10 halo in the early stage of evolution ---
conditions under which the bar instability is actually accelerated. If the increase in the
fraction of chaotic orbits results in a larger velocity dispersion, then the disc becomes
`hotter.'

To verify this, we have measured the radial dispersion velocities, $\sigma_{\rm R}$, in the disc at 
two different times, namely, at $t=0.5$\,Gyr, before the bar instability sets in, and at $t=5$\,Gyr, 
just after the buckling and the spinup
(Fig.\,\ref{fig:P90_sigmaR_Q}). The disc at $t=0.5$\,Gyr is `colder,' and its radial
dispersion velocities are lower. What is more important is that this can be noticed
also by measuring the Toomre's $Q=\kappa\sigma_{\rm R}/3.36 G\Sigma$ parameter, where $\kappa$ and $\Sigma$
are the epicyclic frequency and surface density in the disc, respectively. The condition $Q > 1$ kills the 
axisymmetric 
instabilities in the disc, and $Q > 2-2.5$ damps the non-axisymmetric instabilities \citep[e.g.,][]{binn08}. 
Because, after the stellar bar dissolution, $Q>2$ everywhere outside the inner kpc, the disk in P90Q10 
indeed is too hot to form a bar after buckling.

To further lend support that it is the increased velocity dispersion in the disc that prevents the bar 
from reforming after buckling, we have performed the following numerical test.
We have replaced the spinning halo in P90Q10
model at $t=5$\,Gyr by the spinning halo of P90Q10 model at $t=0$. As Figure\,\ref{fig:A2P90sigma_test} 
demonstrates, the bar instability is completely suppressed when the
disc is immersed in this halo, in a sharp difference with the {\it same} halo at $t=0$. 

So, the combination of spinning halo and buckling are responsible for damping the bar. This explains
why the bar dissolved at 4.5\,Gyr, and only slowly decayed at 8\,Gyr. The stellar orbits have escaped
the dissolved bar at the former time, while remain confined at the latter one. 

\begin{figure}
\centerline{
\includegraphics[width=0.47\textwidth,angle=0] {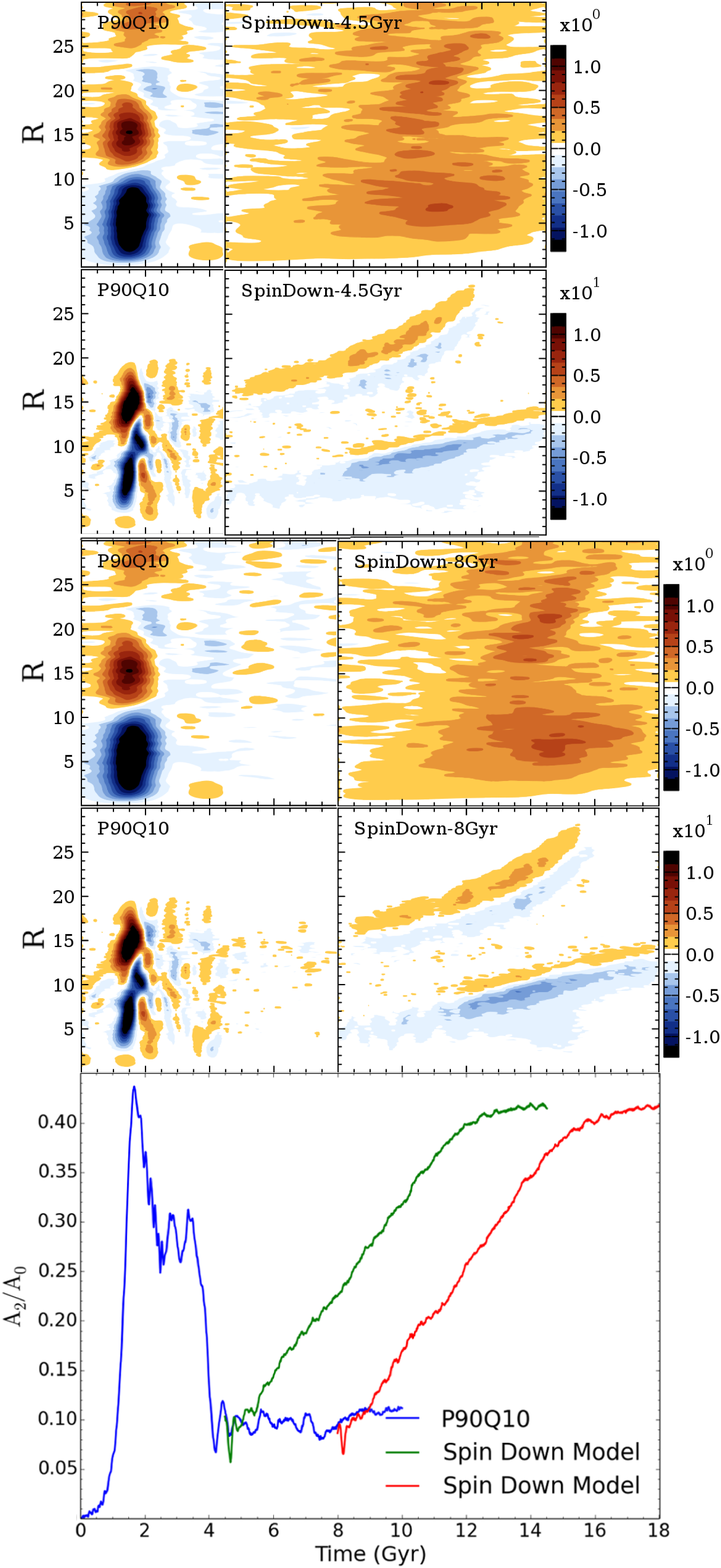}
}
\caption{Spindown of DM halo for P90Q10 model to P00Q10 experiment at  $t=4.5$\,Gyr and 8\,Gyr. 
Same as Figure\,\ref{fig:Jdotmap_spinup}, but for P90Q10 spindown to P00Q10.  
}
\label{fig:Jdotmap_spindown}
\end{figure}

\subsection{Spindown DM halo and bar triggering}
\label{sec:spindown}

The halo spindown tests confirm our reasoning. Figure\,\ref{fig:Jdotmap_spindown}
demonstrates the outcome of the spindown of the DM halo in P90Q10 model to $\lambda=0$
at two different times, $t=4.5$\,Gyr and 8\,Gyr. In both cases the bar instability
sets in and a strong bar develops, exactly as in P00Q10 model, i.e., both test models 
reach the same peak amplitude, matching the value reached by the standard model, P00Q10.

After the spindown, the halo became active in absorbing the disc angular momentum, as
displayed in the maps of angular momentum transfer. At the same time, the disc started
to emit its $J$ from the ILR and showed some absorption around the OLR. This behavior 
clearly demonstrates the  effect of the halo spin on the bar strength.

\subsection{Observational corollaries of bar evolution in spinning DM haloes}
\label{sec:observe}

A long list of observational implications follow from our main result --- modified stellar bar evolution 
with increasing DM halo spin. In this work we touch only a few of these corollaries. 

In order to estimate the importance of this effect, one should account for
the distribution of haloes with $\lambda$. Numerical simulations exhibit a lognormal distribution of
haloes with $\lambda$, with the average of $\bar{\lambda}\sim 0.035-0.04$ 
\citep[e.g.,][]{bull01,hetz06,knebe10}.

\begin{figure*}
\centerline{
\includegraphics[width=0.8\textwidth,angle=0] {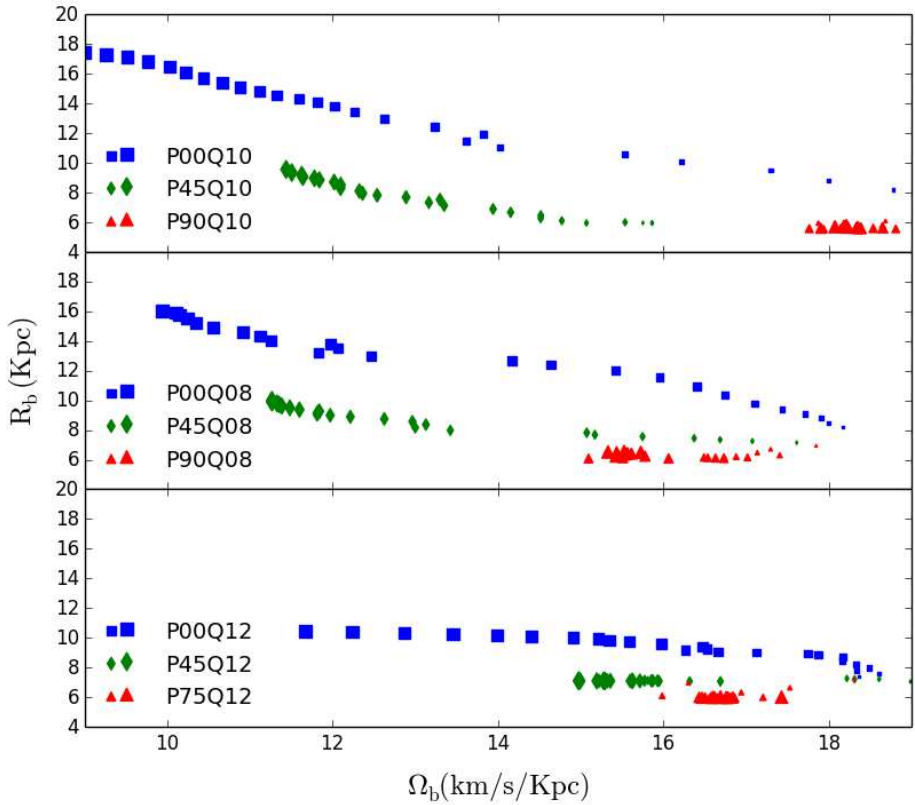}
}
\caption{Evolution of the bar length, $R_{\rm b}$, versus bar pattern speed, $\Omega_{\rm b}$
for spherical (top), oblate (middle) and prolate (bottom) frames. All models are for $\lambda=0$,
0.045 and 0.09. The time direction is given by gradually increasing squares.
}
\label{fig:barL_Omegab}
\end{figure*}

Bars brake against DM haloes as they tumble, which is accompanied by angular momentum transfer
from disc to the DM. As we have discussed earlier, this process involves both resonant and non-resonant 
$J$-transfer. During  this process, bars grow in size. Thus the bar growth and $J$-transfer
are highly correlated.
Figure\,\ref{fig:barL_Omegab} shows a substantially differing evolution of $R_{\rm b}-\Omega_{\rm b}$
correlation along the $\lambda$ sequence and for various shapes of DM haloes. All models have been run
for the same period of time, but occupy different parts of this diagram. Namely, the high $\lambda$
models cluster at high $\Omega_{\rm b}$, especially the prolate models.

The most interesting result is the variation of the final pattern speed of the bars with $\lambda$. The
initial pattern speed in all models is nearly identical. But the final pattern speed has decreased.
The value of this decrease varies from a factor of $\sim 2$ (for $\lambda=0$ models) to just $\sim 5-20\%$ 
(for $\lambda\sim 0.09$ models) below the initial one. 
In fact as we have shown earlier, for a timescale of a few Gyr, bars in the intermediate and higher
$\lambda$ range do not brake at all. These bars, therefore, are genuinely fast bars (not in the sense of
their size compared to the CR radius, which is addressed below).

Next, it has been determined that the ratio $R_{\rm CR}/R_{\rm b}=1.2\pm 0.2$ is a reliable 
indicator for the appearance of offset dust lanes in barred galaxies, which represent the standing shocks
in the gas flow of fast bars. The lower value
comes from the bars being limited by their extent to the CR --- orbits beyond the CR are oriented 
perpendicular to the bar major axis and so cannot support its figure. The upper limit is the 
result found by \citet{atha92a} in 2-D numerical simulations, and represents the slow bars. For 
larger values of $R_{\rm CR}/R_{\rm b}$, the bars are 
substantially shorter of their CR radius and the dust lanes disappear, as a result of the modified gas 
flow. 
 
We find that bars residing within DM haloes with $\lambda > 0.035$ exhibit $R_{\rm CR}/R_{\rm b}$ ratios which 
lie well outside the parameter space provided above, which accommodates the dust lanes.
This is a substantial fraction of haloes, and can accommodate in excess of 50\% of barred discs, which, 
based on the lognormal distribution of $\lambda$, should not exhibit offset dust lanes.
Table\,\ref{table:fast_slow} confirms that the cutoff in halo spin represents well the two groups of bars.
No dependence on the halo shape has been detected.

The tidal torques theory (TTT) distinguishes between the linear phase, when the haloes acquire 
their $\lambda$
and the nonlinear phase. Whether $\lambda$ grows during the later stage is a matter of an ongoing debate
\citep[e.g.,][]{shlo13}. The detailed analysis of spin evolution during mergers has shown
that for a limited time period $\lambda$ increases, then, after relaxation of merger products, it
decreases to the pre-merger value. This has been demonstrated in Figure\,15 of \citet{roma07},
where $\Delta\lambda\sim \pm 0.02-0.03$ [see also \citet{hetz06}]. The typical time of this relaxation for 
massive haloes is $\sim 1-2$\,Gyr. 

\begin{table}
\centering
\caption{Average ratios $R_{\rm CR}/R_{\rm b}$ for various halo shapes and $\lambda$}
\begin{tabular}{l r || l r}
\\
\hline
Halo Shape & Bars regrow       & Bars do not regrow \\
           & $\lambda<0.035$   & $\lambda>0.035$    \\
\hline
Spherical  &     $1.26\pm 0.02$& $2.31\pm 0.15$   \\
Oblate     &     $1.24\pm 0.02$& $2.21\pm 0.15$   \\
Prolate    &     $1.28\pm 0.02$& $2.07\pm 0.15$   \\
\hline
\end{tabular}
\label{table:fast_slow}
\end{table}

This timescale should be compared to the time scale of decay/increase of the bar amplitude discussed
in \S\S\ref{sec:spinup} and \ref{sec:spindown}, which appears to be $\sim 0.5-1$\,Gyr. Given such a short time
scale of bar weakening/strengthening, it is entirely
possible that halo mergers can affect the bar evolution, when the stellar disc survives the ordeal.
 
Formation of {\it ansae} in barred discs is still an unsolved issue \citep[e.g.,][]{marti06}. We detect ansae
in our simulations within all halo shapes considered here. They are persistent for discs with stronger
bars, whether growing or slowly decaying. For example, ansae are present in the slowly decaying bar
of Figure\,\ref{fig:Jdotmap_spinup}, after $t=8$\,Gyr. If we ignore the evolution of bars in the pre-buckling 
phase due to its relatively short time scale, we find ansae in spherical haloes up to $\lambda\sim 0.06$,
in oblate haloes up to $\lambda\sim 0.045$, and in the prolate haloes up to $\lambda\sim 0.03$. 
 
Finally, the peanut/boxy bulges are the direct outcome of the vertical buckling instability in stellar bars.
Moreover, they grow in tandem with the bar growth, as shown by \citet{marti06}. The general trend we observe 
is that the low $\lambda$
models exhibit smaller bulges, irrespective of the halo shape. An additional trend, that has been noticed
already by \citet{long14}, is related to the halo shape which changes from boxy/X-shape in low $\lambda$
models, to boxy in intermediate $\lambda$ haloes, to peanut shapes in higher $\lambda$ haloes. One expects
that the mass and, therefore, the luminosity of these bulges will decrease along the $\lambda$
sequence. We defer this analysis to a later publication.
 
\section{Conclusions}
\label{sec:conc}

To summarize, we have performed a detailed high-resolution study of stellar bar evolution in 
spinning DM haloes, in the range of $\lambda\sim 0-0.09$. We confirm the accelerated bar
instability with increasing $\lambda$, as reported previously, and extend these results to
oblate and prolate haloes. 

Furthermore, we find that secular evolution of stellar bars in spinning haloes results in
damping of their amplitudes along the $\lambda$ sequence. This leads to a decreased transfer
of angular momentum between the disc and its parent halo, and to leveling off the bar
pattern speed. Bars within haloes with larger $\lambda$ have difficulty to re-grow after
a buckling instability. For larger $\lambda$, the bars essentially dissolve, leaving
a weak oval distortion.  

While spinning DM haloes have difficulty to absorb additional angular momentum, it is the
combination of $\lambda$ and the vertical buckling instability of stellar bars that
has a dramatic effect on the bar amplitude, leading to its additional drop and bar dissolution.
The stellar orbits being confined by the bar de-correlate as a result of its dissolution, 
leaving a `hot' disc behind with large radial dispersion velocities.

Damping bars during their secular evolution leads to shorter (slow) bars with 
$R_{\rm CR}/R_{\rm b} > 1.4$, for $\lambda > 0.03$, in contrast to longer (fast) bars in low 
spin DM haloes. 

Although our simulations do not include the gas component, we expect it to have a minor role
in this effect, because the gas is difficult to lock in the resonance due to dissipation.
Yet, the gas can act as to weaken the dynamical instabilities, such as the bar instability
in the vertical buckling in the bar, as noted by \citet{bere98}.

Broad observational corollaries follow from this effect, of which we have mentioned only a few:
$R_{\rm b}-\Omega_{\rm b}$ correlation dependence on the $\lambda$ sequence; absence of
the offset dust lanes in a substantial fraction of barred discs, triggering and damping of
stellar bars in galaxy mergers not by direct tidal torques but by affecting the halo spin;
ansae preference for barred discs in low-$\lambda$ halos; and the shapes of peanut/boxy
bulges, their masses and luminosities.

Hence, stellar bar evolution is substantially more complex when cosmological spin
is taken into account. The central issue is that this evolution demonstrates that 
bars can be destroyed by internal processes in disc-halo systems, or with the help of
external processes, and challenges the present paradigm that stellar bars are resilient entities.

\section*{Acknowledgements}
We thank Phil Hopkins for providing us with the current version of GIZMO.
We are grateful to Sergey Rodionov for help with numerical aspects of initial conditions, and
to Alar Toomre and Scott Tremaine for illuminating discussions.
We thank our colleagues, Jun-Hwan Choi, Yang Luo, Emilio 
Romano-Diaz, Jorge Villa-Vargas and Kentaro Nagamine for help with numerical issues.
This work has been partially supported by the HST/STScI Theory grant 
AR-14584, and by JSPS KAKENHI grant \#16H02163 (to I.S.). 
I.S. is grateful for support from International Joint Research 
Promotion Program at Osaka University. The STScI is 
operated by the AURA, Inc., under NASA contract NAS5-26555. Simulations have been
performed on the University of Kentucky DLX Cluster and using a generous allocation 
on the XSEDE machines to I.S. We thank Vikram Gazula for help with software installation
on the DLX.


\end{document}